\documentclass{article}
\usepackage{spconf,amsmath,graphicx}
\usepackage[noend]{algpseudocode}
\usepackage{subcaption}
\usepackage{soul}
\usepackage{booktabs} 

\usepackage{enumitem}
\usepackage{amssymb}
\usepackage{algorithm}
\usepackage[dvipsnames]{xcolor}

\title{Noise- and Outlier-Resistant Tomographic Reconstruction under Unknown Viewing Parameters}
\name{Ritwick Chaudhry$^{1}\dagger$ \qquad Arunabh Ghosh$^{2}\dagger$ \qquad Ajit Rajwade$^{3}$\sthanks{*AR thanks IIT-B Seed Grant 14IRCCSG012} \thanks{$\dagger$ These authors contributed equally to this work.}}

\address{$^{1}$ Adobe Research \\
         $^{2}$ Dept. of EE, IIT Bombay \\
         $^{3}$ Dept. of CSE, IIT Bombay \\
         rchaudhr@adobe.com; arunabhghosh@iitb.ac.in; ajitvr@cse.iitb.ac.in}
\begin{document}
\ninept
\maketitle
\begin{abstract}
In this paper, we present an algorithm for effectively reconstructing an object from a set of its tomographic projections without any knowledge of the viewing directions or any prior structural information, in the presence of pathological amounts of noise, unknown shifts in the projections, and outliers. We introduce a novel statistically motivated pipeline of first processing the projections, then obtaining an initial estimate for the orientations and the shifts, and eventually performing a refinement procedure to obtain the final reconstruction. Even in the presence of high noise variance (up to $50\%$ of the average value of the (noiseless) projections) and presence of outliers, we are able to reconstruct the object successfully. We also provide interesting empirical comparisons of our method with popular sparsity-based optimization procedures that have been used earlier for image reconstruction tasks.
\end{abstract}
\section{Introduction}
\label{sec:intro}
Reconstructing the structure of an object from its tomographic projections (hereafter referred to as `projections') is a fundamental research problem that arises in diverse fields, such as medical imaging \cite{Fessler2000}, cryogenic  electron microscopy (`Cryo-EM') \cite{frank2006three}, insect tomography \cite{Genise1995ApplicationTraces} and imaging of live, transparent objects \cite{Levis2018}. If the viewing orientations are known \textit{a priori}, standard algorithms such as Filtered Backprojection (FBP) \cite{Feldkamp1984PracticalAlgorithm} can be used to reconstruct the image. However, there are many scenarios where the viewing orientations are unknown. One such example is Cryo-EM where the objective is to determine the structure of a macro-molecule from its projections, which essentially appear in various unknown orientations \cite{frank2006three,Sigworth2016PrinciplesProcessing}. Other examples include insect imaging \cite{Genise1995ApplicationTraces}, transparent live object imaging \cite{Levis2018}, or computed tomography with perturbed geometrical parameters.

\textbf{Prior Work:} In recent times, significant research has emerged in the field of `tomography under unknown viewing parameters'. Much of this research belongs to one of the following two categories: (1) Machine learning based approaches \cite{Basu2000FeasibilityAngles, Fang2011SLLE:Tomography, Coifman2008}, which are based on assumptions on the distribution of the unknown parameters, typically the uniform distribution. However, it might be more likely for a given structure to have specific orientations \cite{Sigworth2016PrinciplesProcessing} and thus would yield the aforementioned assumption unfounded. (2) Moment-based approaches, which uses the Helgason Ludwig Consistency Conditions (HLCC) \cite{Natterer2001TheTomography} to estimate the unknown projection orientations. It has been proved that in the case where the projections are noiseless, a unique solution exists for the equations \cite{Basu2000UniquenessAngles} under some weak assumptions. However, as per our empirical observations, a direct application of the algorithm based on HLCC yields inferior results in practical applications which often bring with them a whole new set of challenges such as (i) severe levels of noise in the projections, (ii) unknown or inaccurately known (albeit small) shifts in the projections, and (iii) presence of outliers. For example, in Cryo-EM, biological specimens are extremely sensitive, so they must be imaged with low-dose electron beams which leads to extremely high amounts of noise in the projections. Moreover, samples of the same biological specimen that are acquired on a single slide may often not be identical due to contaminants such as ice particles, thus adding an extra layer of complexity due to outliers \cite{Huang2016}. The small shifts in the projections are the result of a pre-processing procedure called `particle picking' \cite{Sigworth2016PrinciplesProcessing}.

\textbf{Contributions:} To this end, we present a robust algorithm that can successfully determine the structure of the object from its projections taken in unknown orientations despite the presence of noise, outliers, and unknown shifts without expert intervention or prior structural information. We do not require nor assume any distribution of the projection orientations and thus overcome the limitations in the machine learning based approaches. By demonstrating the ability of our algorithm to perform under high levels of noise and outliers, we signify the effectiveness of this algorithm in practical situations such as Cryo-EM.

In this paper, we work exclusively with 2D images (and hence simulated 1D parallel-beam projections) for reconstruction, even though actual objects are 3D. This follows previous work in the image processing community which has studied the 2D variant of this problem extensively \cite{Basu2000FeasibilityAngles,Singer2013Two-DimensionalDirections}. Nonetheless, the underlying principles remain the same, and the computational problem for reconstructing 2D images remains extremely challenging. Although the focus of this paper is cryo-EM, our algorithm is broadly applicable to other tomography applications with unknown viewing parameters.

\section{Algorithm Description}
\label{sec:format}
\subsection{Overview}
Here, we briefly describe the flow of the proposed algorithm before presenting finer details. We start by systematically clustering the projections and removing the outliers using a robust statistical approach described in Sec. \ref{subsec:robust_clustering} and Sec. \ref{subsec:class1}. Next, we pre-process the projections (the ones not removed in the previous step) to obtain a representative set of less noisy projections using a combination of averaging, principal component analysis (PCA) and classical Wiener filtering described in Sec. \ref{subsec:averaging} and Sec. \ref{subsec:denoising}. In Sec. \ref{subsec:moment_approach} we obtain an initial estimate of the orientations and the shifts of the representative projections using a moment-based approach. Finally, in Sec. \ref{subsec:opt} we optimize for the structure of the unknown object along with a refinement of the viewing parameters (i.e., the angles and the shifts).

\subsection{Robust Clustering}
\label{subsec:robust_clustering}
Typically in the case of Cryo-EM, projections are extremely noisy. Moreover, there may even be a fair number of completely erroneous projections due to the presence of foreign objects and ice particles \cite{Huang2016}. We henceforth refer to these completely erroneous projections as `\textit{outliers of Class 1}'. To combat these problems, we seek to cluster the large number of available projections into a small number of classes, based on orientation and structural similarity. The aim is to produce a representative set of projections that will be significantly less noisy, while simultaneously detecting and rejecting outliers of Class 1. We use the K-means algorithm \cite{Lloyd1982LeastPCM} to cluster the large number of projections into a much smaller number of clusters $K_c$. The distance metric used is the $\ell_r$ quasi-norm ($0 < r \leq 1$) and therefore the cluster centroid is expected to be robust to outliers. The objective function that is minimized is as follows:
\begin{equation}
(L_{centroid})(\{\xi_j\}_{j=1}^{K_c}) = \sum_{j=1}^{K_c}\sum_{p_i \in \chi_j} \|p_i - \xi_j\|_r,
\end{equation}
where there are $K_c$ clusters, $\chi_j$ represents the $j^{\textrm{th}}$ cluster and $\xi_j$ represents the $j^{\textrm{th}}$ cluster centroid. The number $K_c$ should be sufficiently large to capture the characteristics of the projection dataset as well as small enough to fulfill the main motivation of clustering. In our experiments, $K/100$ clusters, where $K$ denotes the total number of projections, was sufficient for good results across datasets. To ensure the robustness of the cluster center to outliers, the value of $r$ is chosen to be 1, and so the cluster centroid is simply the element-wise median of the points belonging to the cluster.

\subsection{Removal of Class 1 Outliers}
\label{subsec:class1}
After clustering, we remove $f\%$ of the projections based on their $\ell_2$ distance from the closest cluster centroid. It is likely that since a completely erroneous projection is located far away from the other projections, a cluster will not be formed close to it. Therefore, removing $f\%$ of projections that are placed farthest from any cluster centroid will remove the Class 1 outliers. A reasonable estimate of $f$ can be provided by a biologist upon eye-balling the micrograph (i.e., an image containing projections of several identical copies of an object, including outliers), and usually, a moderate over-estimate of $f$ is not a problem.

\subsection{Averaging to form a single cluster}
\label{subsec:averaging}
After removal of the Class 1 outliers, we define the processed projection $\tilde{p}_j$ (for cluster index $j$) within each cluster to be the average of all the projections assigned to that cluster, and which were not discarded by the previous step. This is mathematically represented as follows:
\begin{equation}
\tilde{p}_j = \dfrac{\sum_{p_i \in \chi_j} p_i(1-\mathcal{I}_j(p_i))}{\sum_{p_i \in \chi_j} (1-\mathcal{I}_j(p_i))}
\end{equation}
where $\mathcal{I}_j(p_i) = 1$, when the $i^{\textrm{th}}$ projection belongs to the $j^{\textrm{th}}$ cluster is discarded, and 0 otherwise.

\subsection{Patch-Based Denoising}
\label{subsec:denoising}
The processed cluster centers $\{\tilde{p}_j\}_{j=1}^{K_c}$ as obtained in the previous step are significantly less noisy. The residual noise is removed by passing the cluster centers ($\{\tilde{p}_j\}_{j=1}^{K_c}$) through a patch-based PCA denoising algorithm adapted from \cite{MuresanAdaptiveDenoising} (see supplementary material for more details). Hereafter, we use the symbol $\tilde{q}_i$ to refer to the denoised version of the cluster center $\tilde{p}_i$.

\subsection{Initialization of the orientations and shifts using Helgason Ludwig Consistency Conditions (HLCC)}
\label{subsec:moment_approach}
Determining the orientations of the projections and correcting the unknown shifts is a highly non-convex optimization problem. Therefore, we leverage the information available in the image moments and projection moments to obtain an initial estimate of the shifts and the orientations simultaneously. The HLCC \cite{Natterer2001TheTomography} gives us a relationship between the geometric moments of the underlying image $w(x,y)$ and those of its unshifted projections at any angle.

\textbf{HLCC Theorem:} The moments of order $p,q$ of the image $w(x,y)$ are given by $v_{p, q} = \int_{-\infty}^{\infty}\int_{-\infty}^{\infty} w(x,y) x^p y^q dx dy$. The $n^{\textrm{th}}$ order moment of the projection $g(\rho, \theta) \triangleq \int_{-\infty}^{\infty} w(x,y) \delta(\rho - x \cos \theta - y \sin \theta) dx dy$ is given by $
m_{\theta}^{(n)} = \int_{-\infty}^{\infty} g(\rho, \theta) \rho^n d\rho$. If the projection is shifted by $s_i$ to give a projection $g(\rho, \theta, s_i)$, its $n^{th}$ order moment after reverse shifting by an amount $s_k$ can be written as $m_{\theta, s_k}^{(n)} = \int_{-\infty}^{\infty} \mathcal{S} \{g(\rho, \theta, s_i), s_k\} \rho^n d\rho$, where $\mathcal{S} \{., s_k\}$ denotes the reverse shift operation by $s_k$. The above evaluates to the same quantity as $m_{\theta}^{(n)}$ if $s_k = s_i$. That is,
\begin{equation}
m_{\theta, s_i}^{(n)} = \int_{-\infty}^{\infty} \mathcal{S} \{g(\rho, \theta, s_i), s_k\} \rho^n d\rho = m_{\theta}^{(n)}.
\end{equation}
The HLCC give a relationship between $m_{\theta, s_i}^{(n)}$ and $v_{p, q}$, which is defined as
\begin{equation}
m_{\theta, s_i}^{(n)} = \sum_{j=0}^{n} \binom{n}{j} (\cos\theta)^{n-j}(\sin \theta)^j v_{n-j, j}.
\label{eq:hlcc}
\end{equation}
Since, in practice, the projections are noisy, Eqn. \ref{eq:hlcc} will not be satisfied exactly. Instead we define an energy function as follows:
\begin{equation}
E(\{\theta_i\}, \textbf{v}, \{s_i\}) = \sum_{n=0}^k \sum_{i=1}^{K_c} \Bigg ( m_{\theta_i, s_i}^{(n)} - \sum_{j=0}^n A_{i, j}^{(n)} v_{n-j, j} \Bigg )^2.
\label{eq:hlcc_opt}
\end{equation}
Note that in this equation, the moments $m_{\theta_i, s_i}^{(n)}$ correspond to those of the $i^{th}$ cluster center $\tilde{q}_i$ (post-denoising) where $1 \leq i \leq K_c$. Also, $k$ denotes the highest order moment to be considered. In practice, a value of $k = 7$ suffices for all cases. A greater value significantly increases computational time without leading to a discernible increase in gain.
By minimizing this energy function, we derive an initial estimate of the angles and the shifts using an iterative coordinate descent strategy as implemented in \cite{Malhotra2016TomographicRelationships}. A small number of multi-starts (around 10), each with a different random initialization of the pose parameters, helped further combat the non-convexity of the objective function $E(\{\theta_i\}, \textbf{v}, \{s_i\})$. 

\subsection{Optimization strategy to obtain the structure of the object}
\label{subsec:opt}
The estimate provided by the procedure in the previous section, although good, needs to be further refined by a procedure which takes into account the characteristics of the entire original object. For this, we consider the following optimization problem: 
\begin{equation}
\mathcal{M} (\{\theta_i\}, \{s_i\}, w) = \sum_{i=1}^{K_c} \|\tilde{q}_{i, s_i} - \mathcal{R}_{\theta_i} (w)\|^2_2,
\label{eq:fbp_opt}
\end{equation}
where $\tilde{q}_{i, s_i}$ represents a version of $\tilde{q}_i$ shifted by $s_i$, $\mathcal{R}_{\theta_i}$ is the Radon operator at angle $\theta_i$ and $w$ denotes the image to be reconstructed. We solve this in an alternating way, simultaneously refining the estimates of the viewing parameters as well as determining the underlying object structure. First, the structure is estimated using a gradient descent step, which effectively makes use of the FBP algorithm. Next, the orientation and the shift of each projection is estimated using an independent single-dimensional brute-force search. As seen in Sec. \ref{sec:results}, this procedure can determine the structure of the unknown object, while simultaneously refining the viewing parameters even under highly adversarial conditions.

\textbf{Sparsity-based approach:} We also attempted to refine the orientations and shift estimates, along with a refinement of the image $w$ using a sparsity-based optimization technique due to the several promising results delivered by such techniques in the field of compressed sensing \cite{Wang2010}. We considered the following optimization problem: 
\begin{equation}
\mathcal{L} (\{\theta_i\}, \beta, \{s_i\}) = \sum_{i=1}^{K_c} \|\tilde{q}_{i, s_i} - \mathcal{R}_{\theta_i} (U\beta)\|^2_2 + \lambda_1\|\beta\|_1.
\end{equation}
Here $\{\theta_i \}_{i=1}^{K_c},\{s_i \}_{i=1}^{K_c}$ are the $K_c$ unknown angles and shifts for the cluster centers (i.e. $\{\tilde{q}\}_{i=1}^{K_c}$), $\tilde{q}_{i, s_i}$ denotes $\tilde{q}_i$ shifted by $s_i$, $U$ denotes the inverse discrete cosine transform (DCT) operator or any other sparsifying operator, and $\beta$ is the vector of DCT or other transform coefficients of the image to be reconstructed. That is, the image is represented as $w = U \beta$, where $\beta$ is a sparse vector of transform coefficients. 
Upon solving this optimization problem in an alternating fashion using the initial estimates provided by the moment-based estimation, the obtained results exhibited severe artifacts, as we show in Section \ref{sec:results}. A slight error in the initial viewing parameters estimates from Eqn. \ref{eq:hlcc_opt} resulted in the non-convergence of the procedure, highlighting its sensitivity to errors in the original estimate. Therefore we abandoned this line of action and used the former FBP-based optimization technique in Eqn. \ref{eq:fbp_opt} which was significantly more resilient to errors in the initial estimates.

\section{Results}
\label{sec:results}
\begin{figure}[h]
\centering
\begin{subfigure}{0.15\textwidth}
\centering
\includegraphics[width=0.6\linewidth]{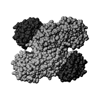}
\caption{Original image}
\end{subfigure}
\begin{subfigure}{0.15\textwidth}
\centering
\includegraphics[width=0.6\linewidth]{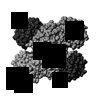}
\caption{Class 2 outlier}
\end{subfigure}
\begin{subfigure}{0.15\textwidth}
\centering
\includegraphics[width=0.6\linewidth]{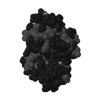} 
\caption{Class 1 outlier}
\end{subfigure}
\caption{The original structure of Holo-Glyceraldehyde-3-Phosphate Dehydrogenase (left). Outlier of class 2 (center). Outlier of class 3 which is a different biological complex - Guanylate Kinase (right).}
\label{fig:outliers}
\end{figure}

In this section, we present a comprehensive set of results obtained using our algorithm on synthetic 2D protein datasets under varying levels of noise, outliers, shifts and different distributions of projection angles. The images used for our experiments were taken from the Database of Macromolecular Movements \cite{Gerstein1998AMotions} and had size $100 \times 100$. A total of $Q = 2 \times 10^4$ projections per image were simulated using angles from $\textrm{Uniform}(0,\pi)$\footnote{Though we considered the Uniform distribution, our algorithm does not rely on this assumption or knowledge of the distribution of the orientations. Later, we also demonstrate our results in the case of a non-uniform distribution of angles.}. A fraction $f_1$ of these projections were outliers of class 1, i.e., they were projections of random images taken from different biological complexes. Another fraction $f_2$ of projections were deliberately generated from a copy of the actual image, but with a small number ($f_3\%$) of pixel values (at randomly selected locations) set to 0. We term the corresponding projections `\textit{outliers of class 2}'. These simulate projections of biological specimens corrupted by overlapping ice particles or minor structural changes. Some sample illustrative images are presented in Fig. \ref{fig:outliers}. All projections were subjected to additive i.i.d. noise from $\mathcal{N}(0,\sigma^2)$, where we assume $\sigma$ to be known in advance, even though there are techniques to estimate it directly from the noisy projections. 

\begin{figure}[h]
\hspace*{-0.3cm}
\centering
\includegraphics[width=0.95\linewidth, height=5cm]{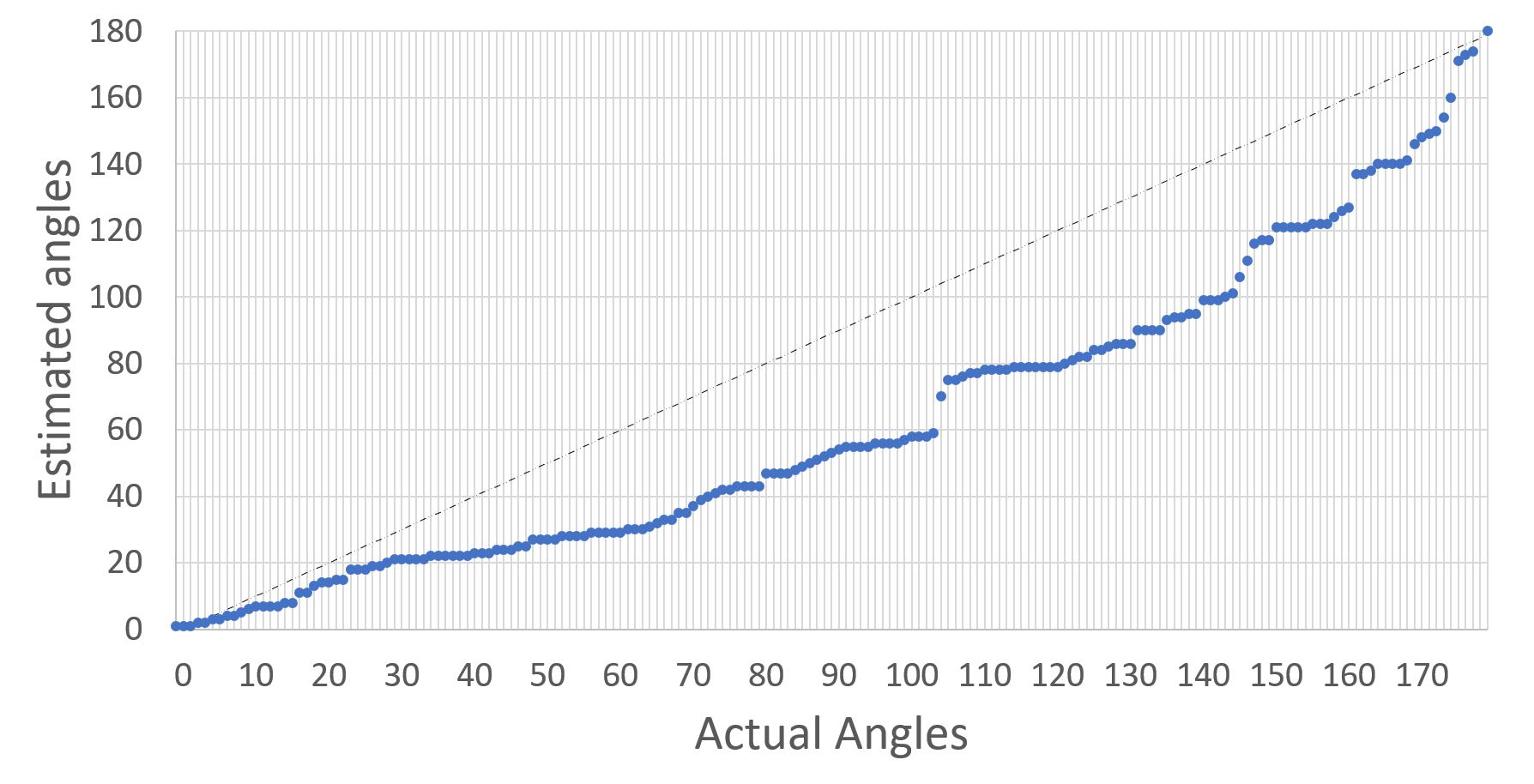} 
\caption{Initial orientation estimates based on Eqn. \ref{eq:hlcc_opt}: $\sigma = 10\%$ Noise, $f_1 = 10\%$ outliers of class 1, $f_2 = 10\%$ outliers of class 2}
\label{fig:moment_orientations}
\end{figure}

\textbf{Performance of the moment-based solver: } The $Q = 2 \times 10^4$ projections are pre-processed to generate $K_c = 180$ less noisy representative projections. These projections are passed to the moments based solver described in Sec. \ref{subsec:moment_approach} which attempts to estimate their respective orientations. Fig. \ref{fig:moment_orientations} shows a scatter plot of the $K_c=180$ estimated projection angles and the corresponding `ground-truth angles' of the cluster centers (the angle of a cluster center is defined as the average of the angles of the projections assigned to that cluster). 

\begin{figure}[h]
\centering
\begin{subfigure}{0.2\textwidth}
\centering
\includegraphics[width=0.42\linewidth]{refined_estimate/cons_2/original_image_1.png} 
\caption{Original Image}
\end{subfigure}
\begin{subfigure}{0.2\textwidth}
\centering
\includegraphics[width=0.43\linewidth]{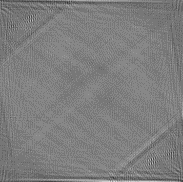}
\caption{Reconstructed Image}
\end{subfigure}
\caption{An example of the failure of sparsity-based optimization. On the left is the original image and on the right is our estimated image.}
\label{fig:csrecon} 
\end{figure}

As seen in Fig. \ref{fig:moment_orientations}, although the moments-based approach provides us with a reasonably good estimate, there is still a need to refine the estimates to obtain high-quality reconstructions. Our initial attempts at employing a sparsity-based optimization framework (implemented using the $\ell_1-\ell_s$ package \cite{Kim2007AnSquares}) failed to yield good quality reconstructions, an example of which is seen in Fig. \ref{fig:csrecon}.

\begin{figure}[h]
\hspace*{-0.3cm}
\centering
\includegraphics[width=0.95\linewidth, height=5cm]{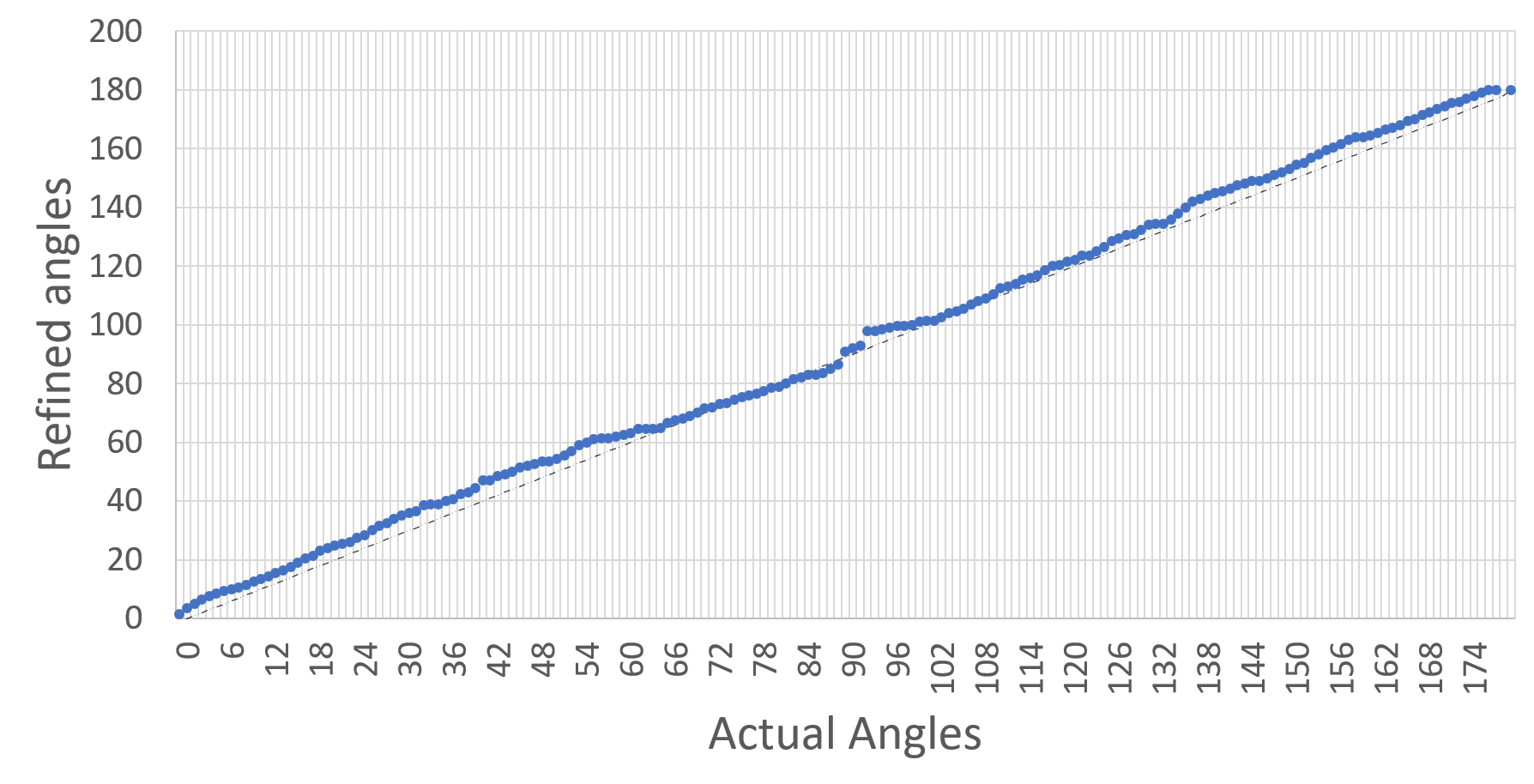} 
\caption{Refined orientation estimates based on Eqn. \ref{eq:fbp_opt}: 10\% Noise, 10\% outliers of class 1, 10\% outliers of class 2}
\label{fig:refined_angles}
\end{figure}

\begin{figure}[H]
\centering
\begin{subfigure}{0.11\textwidth}
\centering
\includegraphics[width=0.74\linewidth]{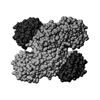}
\caption*{}
\end{subfigure}
\begin{subfigure}{0.11\textwidth}
\centering
\includegraphics[width=0.74\linewidth]{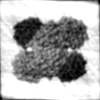}
\caption{16.78\%}
\end{subfigure}
\begin{subfigure}{0.11\textwidth}
\centering
\includegraphics[width=0.74\linewidth]{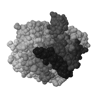}
\caption*{} 
\end{subfigure}
\begin{subfigure}{0.11\textwidth}
\centering
\includegraphics[width=0.74\linewidth]{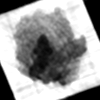}
\caption{17.09\%}
\end{subfigure}

\begin{subfigure}{0.11\textwidth}
\centering
\includegraphics[width=0.74\linewidth]{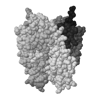}
\caption*{} 
\end{subfigure}
\begin{subfigure}{0.11\textwidth}
\centering
\includegraphics[width=0.74\linewidth]{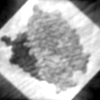}
\caption{18.85\%}
\end{subfigure}
\begin{subfigure}{0.11\textwidth}
\centering
\includegraphics[width=0.74\linewidth]{refined_estimate/cons_2/original_image_1.png}
\caption*{}
\end{subfigure}
\begin{subfigure}{0.11\textwidth}
\centering
\includegraphics[width=0.74\linewidth]{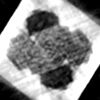}
\caption{24.47\%}
\end{subfigure}
 
\begin{subfigure}{0.11\textwidth}
\centering
\includegraphics[width=0.74\linewidth]{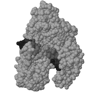}
\caption*{}
\end{subfigure}
\begin{subfigure}{0.11\textwidth}
\centering
\includegraphics[width=0.74\linewidth]{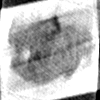}
\caption{26.44 \%}
\end{subfigure}
\begin{subfigure}{0.11\textwidth}
\centering
\includegraphics[width=0.74\linewidth]{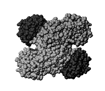} 
\caption*{}
\end{subfigure}
\begin{subfigure}{0.11\textwidth}
\centering
\includegraphics[width=0.74\linewidth]{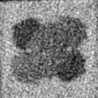}
\caption{32.31\%}
\end{subfigure}
\caption{Each row demonstrates two reconstructions under the following conditions; Top row: 0\% Noise and no outliers of class 1 or class 2; Second row: 10\% Noise, 10\% outliers of class 1 and 2; Third row: 50\% Noise, 5\% outliers of class 1 and 2. In each case, original image is shown on the left and the reconstructed image is shown next to it. The RMSE is written below each reconstruction.}
 \label{fig:reconstruction_gen}

\vspace{0.1cm}
\begin{subfigure}{0.11\textwidth}
\centering
\includegraphics[width=0.74\linewidth]{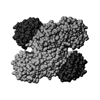}
\caption*{}
\end{subfigure}
\begin{subfigure}{0.11\textwidth}
\centering
\includegraphics[width=0.74\linewidth]{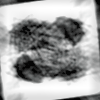} 
\caption{29.07\%}
\end{subfigure}
\begin{subfigure}{0.11\textwidth}
\centering
\includegraphics[width=0.74\linewidth]{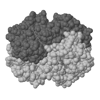}
\caption*{}
\end{subfigure}
\begin{subfigure}{0.11\textwidth}
\centering
\includegraphics[width=0.74\linewidth]{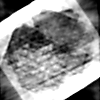}
\caption{30.06\%}
\end{subfigure}
\caption{2 examples of reconstructions in the case the distribution of orientations is non-uniform. The relevant parameters are: 20\% Noise, 5\% outliers of class 1, 5\% outliers of class 2. In each case, original image is shown on the left and the reconstructed image is shown next to it. The RMSE is written below each reconstruction.}
\label{fig:recon_non_uniform} 

\vspace{0.1cm}
\begin{subfigure}{0.11\textwidth}
\centering
\includegraphics[width=0.74\linewidth]{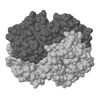} 
\caption*{}
\end{subfigure}
\begin{subfigure}{0.11\textwidth}
\centering
\includegraphics[width=0.74\linewidth]{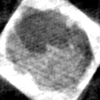}
\caption{20.77\%}
\end{subfigure}
\begin{subfigure}{0.11\textwidth}
\centering
\includegraphics[width=0.74\linewidth]{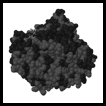} 
\caption*{}
\end{subfigure}
\begin{subfigure}{0.11\textwidth}
\centering
\includegraphics[width=0.74\linewidth]{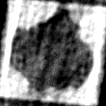}
\caption{20.95\%}
\end{subfigure}
\caption{2 examples of reconstructions in the case where the projections have random unknown shifts upto $\pm 2$. The relevant parameters are: 10\% Noise, 0\% outliers of class 1, 0\% outliers of class 2. In each case, original image is shown on the left and the reconstructed image is shown next to it. The RMSE is written below each reconstruction.}
\label{fig:shift_reconstruction}
\end{figure}

We, therefore, use the more robust FBP based approach to perform a simultaneous refinement of the structure as well the viewing parameters. As shown in Fig. \ref{fig:refined_angles} the error between the refined angles and the ground-truth angles is significantly less when compared to the original estimates (Fig. \ref{fig:moment_orientations}). In Fig. \ref{fig:reconstruction_gen} we exhibit some of the reconstructions achieved by this algorithm under varying levels of noise and outliers. Even in the case of pathological amounts of noise, we are able to estimate the structures to a high degree of accuracy. The error metric used to assess the quality of reconstruction is the Relative Mean Squared Error (RMSE) between the registered reconstruction (reconstruction aligned with the test image) and the test image. The registration is needed since the solution can be obtained only up to a global rotation/translation. The RMSE is defined as $\textrm{RMSE}(w,\hat{w}) = \|w-\hat{w}\|_2/\|w\|_2$, where $\hat{w}$ is the registered reconstructed estimate for $w$. 

\textbf{Experiments with non-uniform orientation distributions:} In this experiment, we consider the following distribution for the projection orientations: 
$\textrm{Uniform}(0,\pi/9) \cup \textrm{Uniform}(2\pi/9,\pi/3) \hspace{0.1cm}\cup
       \textrm{Uniform}(4\pi/9,2\pi/3) \cup \textrm{Uniform}(7\pi/9,8\pi/9) 
$. The reconstructions obtained in Fig. \ref{fig:recon_non_uniform} demonstrates that we do not require nor assume anything about the underlying distribution of orientations.

\textbf{Experiments with unknown shifts: } Even if the projections are afflicted with random translational errors, our algorithm automatically estimates as well as corrects the shifts to obtain accurate reconstructions as shown in Fig. \ref{fig:shift_reconstruction}.

\section{Discussion and Conclusion}
\label{sec:typestyle}
From the results presented in the previous section, we conclude that our algorithm is capable of estimating the original structure from its tomographic projections at unknown angles, even in the case of pathological amounts of noise and a high percentage of outliers and unknown shifts in the projections. Further, our method does not make any assumption on the distribution of the projection angles, a limitation of the prior solutions which often assumes a uniform distribution of angles. An important avenue of future investigation is analyzing the failure of the sparsity-based optimization framework despite the promising results of compressed sensing in many applications. Moreover, we will extend our algorithm to the three-
\linebreak

\noindent
dimensional case and validate it on actual Cryo-EM datasets, insect tomography datasets, and CT reconstructions with patient motion.

\noindent
\textbf{Supplemental material:} For an overall summary of the algorithm and additional results refer to the supplemental material.

\noindent
\textbf{Note:} The authors have submitted a different paper \cite{GhoshHeterogeneity} which exclusively deals with ab initio tomographic reconstruction of \emph{heterogeneous} objects (i.e. objects with multiple structures or `conformations'). The problem of heterogeneity cannot be solved by reconstructing one conformation while considering the projections of other conformations as outliers. This is since our algorithm described here deals with only a small percentage of outliers. On the other hand, unlike the work in this paper, the work in \cite{GhoshHeterogeneity} does not handle outliers of class 1 and class 2.

\newpage
\bibliographystyle{IEEEbib}
\bibliography{main}

\end{document}